\documentclass[prl,aps]{revtex4}
\usepackage{epsfig,rotate}
\usepackage{graphicx}
\newcommand{\vnabla}{{\mbox{\boldmath$\nabla$}}}

\newcommand{\vD}{{\mbox{${\bf D}$}}}

\newcommand{\vA}{{\mbox{${\bf A}$}}}

\newcommand{\vQ}{{\mbox{${\bf Q}$}}}

\newcommand{\rhoqx}{{\mbox{$\rho_{2{\bf Q}_x}$}}}
\newcommand{\rhoqy}{{\mbox{$\rho_{2{\bf Q}_y}$}}}
\newcommand{\rhoqxyn}{{\mbox{$\rho_{{\bf Q}_x-{\bf Q}_y}$}}}

\newcommand{\rhoqxyp}{{\mbox{$\rho_{{\bf Q}_x+{\bf Q}_y}$}}}

\newcommand{\sqxyn}{{\mbox{$S^z_{{\bf Q}_x-{\bf Q}_y}$}}}
\newcommand{\sqxyp}{{\mbox{$S^z_{{\bf Q}_x+{\bf Q}_y}$}}}

\newcommand{\vB}{{\mbox{${\bf B}$}}}

\newcommand{\Dpx}{{\mbox{$\Delta_{{\bf Q}_x}$}}}
\newcommand{\Dpy}{{\mbox{$\Delta_{{\bf Q}_y}$}}}
\newcommand{\Dnx}{{\mbox{$\Delta_{-{\bf Q}_x}$}}}
\newcommand{\Dny}{{\mbox{$\Delta_{-{\bf Q}_y}$}}}
\newcommand{\Dpq}{{\mbox{$\Delta_{{\bf Q}}$}}}
\newcommand{\Dnq}{{\mbox{$\Delta_{-{\bf Q}}$}}}
\newcommand{\Dpqi}{{\mbox{$\Delta_{{\bf Q}_i}$}}}
\newcommand{\Dnqi}{{\mbox{$\Delta_{-{\bf Q}_i}$}}}
\newcommand{\Dpqj}{{\mbox{$\Delta_{{\bf Q}_j}$}}}
\newcommand{\Dnqj}{{\mbox{$\Delta_{-{\bf Q}_j}$}}}

\newcommand{\Dd}{{\mbox{$\Delta_d$}}}
\begin{document}

\title{Dislocations and vortices in pair density wave superconductors}

\author{D.F. Agterberg$^{1,2}$  and H. Tsunetsugu$^{1}$}
\address{$^1$ Institute for Solid State Physics, University of Tokyo, Kashiwa, Chiba 277-8581, Japan}
\address{$^2$ Department of Physics, University of Wisconsin-Milwaukee, Milwaukee, WI 53211}

%\PACS{74.20.De,74.25.Dw }

\begin{abstract}

With the ground breaking work of the Fulde, Ferell, Larkin, and
Ovchinnikov (FFLO), it was realized that superconducting order
can also break translational invariance; leading to a phase in
which the Cooper pairs develop a coherent periodic spatially
oscillating structure.  Such pair density wave (PDW)
superconductivity has become relevant in a diverse range of
systems, including cuprates, organic superconductors, heavy
fermion superconductors, cold atoms, and high density quark
matter. Here we show that, in addition to charge density wave
(CDW) order, there are PDW ground states that induce spin density
wave (SDW) order when there is no applied magnetic field.
Furthermore, we show that PDW phases support topological defects
that combine dislocations in the induced CDW/SDW order with a
fractional vortex in the usual superconducting order. These
defects provide a mechanism for fluctuation driven
non-superconducting CDW/SDW phases and conventional vortices with
CDW/SDW order in the core.

\end{abstract} \maketitle

PDW superconductivity, of the kind originally discussed by FFLO
\cite{lar65,ful64}, is believed to exist in the heavy fermion
superconductor CeCoIn$_5$  \cite{rad03,bia03} and  in the organic
superconductor $\kappa$-(BEDT-TTF)$_2$Cu(NCS)$_2$
\cite{lor07,mat07}. It is also believed be relevant in cold atoms
\cite{miz05,yan05} and in the formation of color
superconductivity in high density quark matter \cite{cas04}.
Also,  PDW superconductivity with quite a different origin has
been found in microscopic theories of correlated electronic
materials \cite{him02,rac07,ali07}. Such order is believed by
some to be competing with conventional $d$-wave superconductivity
in underdoped high temperature cuprate superconductors.
Specifically, PDW order may appear as an alternate ground state
near a hole doping of 1/8, where some sort of charge order is
often observed \cite{ber07,che04}.  Given its relevance, it is
important and useful to address the properties of PDW order from
a phenomenological point of view. That is the primary goal of
this work. In particular, we address two questions that apply to
both FFLO superconductors and the underdoped cuprates: 1) What
symmetries are broken by PDW order? and  2)  What are the
properties of the vortex-like topological defects in the PDW
ordered phases? The answer to the first question reveals that CDW
or SDW order often must accompany the PDW order and provide a
means to identify the PDW phase. The answer to the second
question turns out to be non-trivial. It is clear that there will
be superconducting vortices and these have been studied in the
past (see Ref.~\cite{mat07} for an overview). However, it is also
well known that the periodic order has dislocations as natural
topological defects. This has not been given much consideration
in the context of PDW order. Here we show that the topological
defects of PDW superconductors contain not just vortices and
dislocations, but also combinations of fractional vortices and
fractional dislocations. We argue that these fractional defects
play an important role when considering fluctuations in
two-dimensions (for example, leading to non-superconducting CDW
or SDW phases) and that they play an important role when
considering the physics of the usual superconducting vortices
(for example, leading to the appearance of CDW or SDW order
inside the vortex core). Finally, we address an issue that is
mainly of relevance to the underdoped cuprates. In particular, we
examine the role of competition between PDW order and
translational invariant $d$-wave superconductivity. We show that
this competition preferentially selects PDW phases with CDW
order. Furthermore, in addition to the CDW order stemming from
the PDW order parameter, the coexistence of PDW and $d$-wave
superconductivity leads to either SDW or additional CDW order.

While our main results apply more generally to PDW
superconductors, for concreteness we consider an example motivated
by theoretical and experimental proposals for the underdoped
cuprates that exhibit some type of charge order. In particular, we
consider a PDW superconductor in three dimensional tetragonal
system with a lattice spacing $a$ for the two-dimensional square
lattices. The PDW order is taken to be either commensurate, with a
periodicity of $8a$ \cite{him02,rac07,ber07,che04}, or
incommensurate (this also allows the theory to describe a variety
of FFLO phases found to be stable in two-dimensions \cite{mat07}).
Furthermore, we will take this order to be aligned along the
$\hat{x}$ (taken to be along a two-fold symmetry axis) or
equivalent directions. The PDW order parameter is written as
$\{\Dpx,\Dpy,\Dnx,\Dny\}$ describing PDW order with corresponding
wavevectors $\{\vQ_x,\vQ_y,-\vQ_x,-\vQ_y\}$. The Ginzburg Landau
free energy density is constructed by imposing translational,
gauge, time-reversal, parity, and tetragonal rotation symmetries.
This yields
\begin{eqnarray}
f=&\alpha\sum_{i}|\Dpqi|^2+\beta_1(\sum_{i}\Dpqi|^2)^2 +
\beta_2\sum_{i<j}|\Dpqi|^2|\Dpqj|^2+\beta_3(|\Dpx|^2|\Dnx|^2+|\Dpy|^2|\Dny|^2) \label{free}\\
&+\beta_4[\Dpx\Dnx(\Dpy\Dny)^*+(\Dpx\Dnx)^*\Dpy\Dny]
\nonumber\\
&+\kappa_1\sum_i|\vD \Dpqi|^2+ \kappa_2\sum_i (-1)^i(
|D_x\Dpqi|^2-|D_y\Dpqi|^2)+ \kappa_3\sum_i|D_z\Dpqi|^2
+\frac{1}{2}(\vnabla\times\vA)^2\nonumber
\end{eqnarray}
where $(-1)^i$ is $-1$ for $\vQ_i=\pm\vQ_x$ and $1$  for
$\vQ_i=\pm\vQ_y$, $\vD=-i\vnabla- 2e \vA$, and
$\vB=\vnabla\times\vA$. The difference between the commensurate
and incommensurate cases arises at eighth order. This implies that
sufficiently near the mean field $T_c$, this difference can be
ignored. However, there are situations where this difference can
be important and this will be discussed later (in particular, when
considering vortex-type excitations in a two-dimensional PDW
superconductor).

An important property of the above free energy is that it contains
a $U(1)\times U(1) \times U(1)$ symmetry. It is this feature that
gives rise to vortices and dislocations as well as the appearance
of the combined fractional vortex and fractional dislocations
mentioned above. Below, we will provide a physical picture for the
topological defects of Eq.~\ref{free} (and also for the related
half-flux vortices found in Ref.~\cite{agt07}). Prior to
understanding these defects, it is important to first understand
the possible PDW ground states and also useful to understand the
induced charge density wave (CDW) and spin density wave (SDW)
order that arises.  To complete the latter goal, we include the
free energy for CDW and SDW order and their coupling to the PDW
order. These are also determined by the same symmetry requirements
as above. The contribution to the free energy from the CDW order,
$\rho_{\bf Q}$ is
\begin{equation}
\sum_{i,j}\{\delta_{i,j}\alpha^{\rho}_{{\bf Q}_i}\rho_{{\bf
Q}_i}\rho_{-{\bf Q}_i}+ \alpha^{\rho}_{i,j}\rho_{{\bf Q}_j-{\bf
Q}_i}[\Dpqi(\Dpqj)^*+\Dnqj(\Dnqi)^*].
\end{equation}
For our choice of PDW order, only SDW order with moments oriented
along $\hat{z}$ can be induced.  Denoting the SDW order as
$S^z_{\bf Q}$, the relevant free energy is
\begin{equation}
\sum_{i,j}\{\delta_{i,j}\alpha^{s}_{{\bf Q}_i}S^z_{{\bf
Q}_i}S^{z}_{-{\bf Q}_i}+\alpha^{s}_{i,j}S^{z}_{{\bf Q}_j-{\bf
Q}_i} i[\Dpqi(\Dpqj)^*-\Dnqj(\Dnqi)^*]\}.
\end{equation}
Throughout this work we consider the case that the CDW and SDW
order is induced by the PDW order ($\alpha^{\rho}_{\bf
Q},\alpha^s_{\bf Q}>0$) and thus is representative of the
symmetries broken by the PDW phase transition. In this case they
are given by
\begin{equation}
\rho_{{\bf Q}_i-{\bf
Q}_j}=-\frac{\alpha^{\rho}_{ij}}{2\alpha^{\rho}_{{\bf Q}_i-{\bf
Q}_j}} [\Dpq_i (\Dpq_j)^*+ \Dnq_j (\Dnq_i)^*]\label{cdw}
\end{equation}
\begin{equation}
S^z_{{\bf Q}_i-{\bf Q}_j}=-\frac{\alpha^s_{ij}}{2\alpha^s_{{\bf
Q}_i-{\bf Q}_j}} i[\Dpq_i (\Dpq_j)^*- \Dnq_j
(\Dnq_i)^*]\label{sdw}.
\end{equation}
Eqs. \ref{cdw} and \ref{sdw} reveal that the phase difference
between two different components $\Dpq$ can be interpreted as the
phase of either the CDW or SDW order. In general, it is possible
that there is also an intrinsic CDW, or SDW order (for which
$\alpha^{\rho}_{\bf Q}\le0$ or $\alpha^{s}_{\bf Q}\le0$ is
possible). This is also of interest but is not considered in this
work.

% For future reference the free energies of the phases are given as:
% Phase 1: b1
% Phase 2: b1+b2/2
% Phase 3: b1+b2/2+b3/4
% Phase 4: b1+(3/4)b2+b3/8+b4/8
% Phase 5: b1+(3/4)b2+b3/8-b4/8
% It is possible to plot the stability regimes on a plot of b2,b3
\begin{table}
\begin{tabular}{|c|c|c|c|c|c|}
  \hline
  Phase & $(\Dpx,\Dpy,\Dnx,\Dny)$& Stability & CDW modes & SDW modes &  Flux \\
  \hline
  1& $(e^{i\phi_1},0,0,0)$ & $\beta_2>0,\beta_2+\beta_3>0$  &none & none & $\Phi_0$\\
  &&$3\beta_2+\beta_3-|\beta_4|>0$&&&\\
  2 & $(e^{i\phi_1},e^{i\phi_2},0,0)$& $\beta_2<0$ & $\rhoqxyn$ & $\sqxyn$& $\Phi_0/2$\\
  &&$\beta_2+\beta_3-|\beta_4|>0$&&&\\
  3 & $(e^{i\phi_1},0,e^{i\phi_2},0)$& $\beta_2+\beta_3<0$ & $\rhoqx$ & none & $\Phi_0/2$\\
  &&$\beta_2-\beta_3-|\beta_4|>0$&&&\\
  4 & $(e^{i\phi_1},e^{i\phi_2},e^{i\phi_3},e^{i[\phi_1+\phi_3-\phi_2]})$&$\beta_4<0,\beta_2+|\beta_3|<|\beta_4|$ & $\rhoqx,\rhoqy$
  & none & $\Phi_0/2$  \\
  &&$3\beta_2+\beta_3-|\beta_4|<0$&$\rhoqxyn,\rhoqxyp$&&($\Phi_0/4$)\\
  5 & $(e^{i\phi_1},ie^{i\phi_2},e^{i\phi_3},ie^{i[\phi_1+\phi_3-\phi_2]})$& $\beta_4>0,\beta_2+|\beta_3|<|\beta_4|$ & $\rhoqx,\rhoqy$ & $\sqxyn,\sqxyp$
  & $\Phi_0/2$ \\
  &&$3\beta_2+\beta_3-|\beta_4|<0$&&&($\Phi_0/4$)\\
  \hline
\end{tabular}
\caption{{\bf Properties of PDW Ground States.} All possible PDW
ground states and accompanying CDW and SDW order. The third column
shows the parameter regions for which these phases are stable. In
the fourth and fifth columns other modes can be found by using the
relationships $\rho_{\bf Q}=(\rho_{-{\bf Q}})^*$ and $S^z_{\bf
Q}=(S^z_{-{\bf Q}})^*$. The fifth column shows the minimum flux
contained by a topological defect. In Phases 4 and 5, the
$\Phi_0/4$ defects experience a confinement potential but can
exist at short length scales and may consequently become
physically relevant.} \label{tab1}
\end{table}

Remarkably, it is possible to find all the possible PDW ground
states of the general free energy defined by Eqs.~\ref{free}
(this is not true, for example, for a general phenomenological
theory of superfluid $^3$He \cite{leg75}). These ground states
and the induced CDW and SDW orders are listed in
Table~\ref{tab1}. Note that we have listed the lowest Fourier
components of the induced SDW and CDW order, higher Fourier
harmonics will in general appear but these should be smaller in
magnitude.   In principle, the induced CDW and SDW order can be
measured and therefore used to identify what type of PDW order
appears. One result that has not previously been highlighted is
the existence of PDW phases that break time-reversal symmetry. In
particular, Phases 2 and 5 exhibit this behavior and the broken
time reversal symmetry is readily apparent through the associated
SDW order. At the mean field level, this theory predicts that
superconductivity appears at the same transition temperature as
the SDW and CDW order. However, fluctuations can separate these
transitions and a specific mechanism for this will be discussed
later. It is interesting that related SDW and CDW order has been
observed in a variety of cuprates \cite{ver04,han04,yee99} and a
systematic experimental study of such order in a single material
would prove useful to identify any PDW order parameter.

We now address the nature of the competition between
translational invariant $d$-wave superconductivity and PDW
superconductivity relevant only in the context of the cuprates.
The lowest order coupling between these two order parameters is
given by the following free energy density:
\begin{equation}
f_c=\beta_{c1} \sum_i |\Dd|^2|\Dpqi|^2 +
\beta_{c2}[\Dd^2(\Dpx\Dnx+\Dpy\Dny)^*+(\Dd^2)^*(\Dpx\Dnx+\Dpy\Dny)].
\label{PDWdc}
\end{equation}
The relevant feature of this coupling is that the term with
coefficient $\beta_{c2}$ can always be made negative by an
appropriate choice of phases of the different order parameters.
Consequently, any PDW phase for which this coupling is non-zero
will lower its energy through this coupling. Only Phases 3 and 4
above lead to a non-zero $\beta_{2c}$ coupling, these are the two
phases with PDW induced CDW order. When PDW and $d$-wave
superconductivity co-exist, then the interplay between these two
order parameters will lead to CDW or SDW order that appears in
addition to the CDW order that already exists in Phases 3 or 4.
In particular, if the coefficient $\beta_{c2}>0$ then the
additional order will be SDW order at the wavevectors of the PDW
order; if  $\beta_{c2}<0$, then there will be no induced SDW
order and additional CDW order will appear at the wavevectors of
the PDW order.

%In the broken time-reversal symmetry phases,  the
%interrelationship between PDW, CDW, and SDW order is non-trivial.
%For example in Phase 2, the SDW and CDW order appear at the same
%wavevector. However, the spatial pattern is such that SDW order
%is largest where the CDW order vanishes and vice vera. Phase 5 is
%also intriguing, in this case the CDW and SDW order appear at
%different wavevectors.  For example, if $a\vQ_x=(3\pi/4,0)$ and
%$a\vQ_y=(0,3\pi/4)$. Then there will be CDW order at wavevectors
%$a\vQ=(\pm\pi/2,0)$ and $(0,\pm\pi/2)$. The SDW order will appear
%at wavevectors $a\vQ= (-3\pi/4,\pm3\pi/4)$ and $(\pm
%3\pi/4,3\pi/4)$.

\begin{figure}
\epsfxsize=3.0 in
\rotatebox{270}{\centerline{{\epsfbox{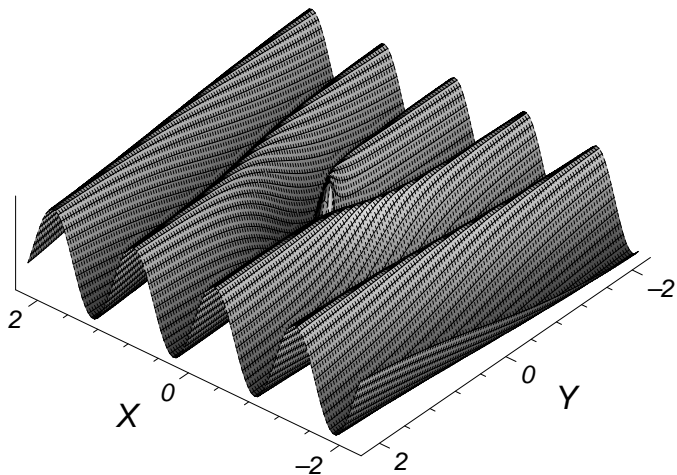}}}}
%%\center{\epsfbox{fig1-2.eps}}
\caption{{\bf Fractional Vortex -
Dislocation.} Dislocation in CDW order accompanying a flux
$\Phi_0/2$ defect in Phase 3. Since the periodicity of the PDW
order is twice that of the CDW order, this corresponds to half a
dislocation in the PDW order.} \label{fig1}
\end{figure}

With the understanding of the possible PDW mean-field ground
states, we now turn to understanding the topological defects.
Single valuedness of the wavefunctions implies that these can be
found by allowing the phases $\phi_1,\phi_2$, or $\phi_3$ in
Table~\ref{tab1} to have an integral multiple of phase winding
$2\pi$ upon encircling the core of the defect. To understand the
energies and the magnetic properties of such defects, it is
useful to consider a London theory. Here we consider Phase 3
explicitly  and state the results for the other phases. Allow
$(\phi_1,\phi_2)$ to have phase windings of $(n,m)$ times $2\pi$
respectively when encircling a defect. Taking
$(\Dpx,\Dpy,\Dnx,\Dny)=\psi(e^{i\phi_1},0,e^{i\phi_2},0)/\sqrt{2}$,
gives the London theory
\begin{equation}
F=\int d^3r \sum_i \{\frac{\rho_{s,i}}{2}[(\partial_i
\phi_1-2eA_i)^2+(\partial_i \phi_2-2eA_i)^2]+\frac{1}{2}B_i^2
\}\label{free2}
\end{equation}
where $\rho_{s,x}=(\kappa_1+\kappa_2)|\psi|^2$,
$\rho_{s,y}=(\kappa_1-\kappa_2)|\psi|^2$, and
$\rho_{s,z}=\kappa_3|\psi|^2$. Closely related London theories
appear in a variety of other contexts, including in a theory
motivated by superconducting UPt$_3$ \cite{zhi95}, in two-gap
superconductors \cite{bab02}, and a description of fractional
vortices (with non-Abelian vortex core states) in chiral
superconductors \cite{chu07}.  Eq.~\ref{free2} yields  a
supercurrent with components $J_i\propto\rho_{s,i}[\partial_i
(\phi_1+\phi_2)-2eA_i]/2$, far from the core of the defect, the
minimum energy configuration has zero current and a contour
integration then implies that the flux enclosed by a $(n,m)$
defect is $(n+m)\Phi_0/2$. Consequently, the simplest defect
encloses half a flux quantum. A similar analysis shows that Phase
1 has only the usual superconducting vortices (the simplest
containing flux $\Phi_0$) while Phases 2,3,4, and 5 all have
defects that contain flux $\Phi_0/2$. Eq.~\ref{free2} also shows
that usual superconducting vortices have finite energy because
the phase winding can be completely screened by the vector
potential. The other defects have an energy that diverges as the
logarithm of the system size.  To help develop a physical picture
of the fractional defects, it is useful to examine the induced
CDW (and SDW) order near a $\Phi_0/2$ defect. Fig.~\ref{fig1}
reveals that near such a defect in Phase 3 there is an edge
dislocation in the CDW order when spatial uniformity of the
defect along the $\hat{z}$ direction is imposed. If the defect is
taken to be spatially uniform along the direction parallel to
$\vQ$, then a screw dislocation appears in the CDW order.  Note
that the CDW order in this phase has half the periodicity of the
PDW order, consequently, a dislocation in the CDW order can be
interpreted as half a dislocation in the PDW order. The
$\Phi_0/2$ defects of the other phases have a similar origin.
This leads to the prediction that a PDW superconducting vortex
containing a fraction of a flux quantum will be pinned to a
dislocation in the accompanying CDW or SDW order. It is
worthwhile pointing out that if $\beta_4=0$ in Eq.~\ref{free},
then Phases 4 and 5 will have $\Phi_0/4$ defects. A nonzero
$\beta_4$ leads to a confinement potential between these
$\Phi_0/4$  defects. However, it is still possible that the
$\Phi_0/4$ defects play a role at short length scales, for
example in determining the vortex core structure of integer flux
superconducting vortices.

%It may be of interest that there exist PDW phases with defects
%that contain fractions smaller than $\Phi_0/2$. For example, in
%tetragonal systems, defects with $\Phi_0/4$ flux are possible in
%a phase described by  $\Delta(e^{i\mbox{\boldmath{\scriptsize$
%q$}}_1\cdot {\bf R}}+e^{i\mbox{\boldmath{\scriptsize$ q$}}_2\cdot
%{\bf R}}+e^{i\mbox{\boldmath{\scriptsize$ q$}}_3\cdot {\bf
%R}}+e^{i\mbox{\boldmath{\scriptsize$ q$}}_4\cdot {\bf R}})$,
%where $\mbox{\boldmath$ q$}_1=(q_x,q_y)$ is along a low-symmetry
%direction and the four $\vq_i$ are such that for each  $\vq_i$,
%$-\vq_i$ is not one of the other three $\vq_j$. Another example
%is in materials with hexagonal symmetry where $\Phi_0/3$  defects
%exist. These defects appear in a phase with order parameter
%$\Delta(e^{i\mbox{\boldmath{\scriptsize$ q$}}_1\cdot {\bf
%R}}+e^{i\mbox{\boldmath{\scriptsize$ q$}}_2\cdot {\bf R}
%}+e^{i\mbox{\boldmath{\scriptsize$ q$}}_3\cdot {\bf R}})$, where
%$\vq_1+\vq_2+\vq_3=0$. Such an order parameter has been argued to
%be stable in mean-field theories of two-dimensional FFLO
%superconductors \cite{mat07}.

We now turn to two physical consequences of the fractional flux
defects. The first is in fluctuation driven vortex and
dislocation physics in two dimensions. In this case, the
fractional flux defects lead to non-superconducting CDW/SDW
phases.  The second consequence is that conventional
superconducting vortices can decay into a bound state of
fractional flux defects.  This provides a mechanism for the
appearance of CDW or SDW order inside a vortex core.

The fact that single flux quanta vortices have short range
interactions  and fractional flux defects have long range
interactions imply that novel vortex related physics can occur in
two dimensions.  This physics differs for the commensurate and
incommensurate cases and discussion of the commensurate case is
left to the next paragraph. The theory for the incommensurate
case follows from the London theory of Eq.~\ref{free2} and
resembles that for two band superconductors considered in
Ref.~\cite{bab04}. To understand the behavior of the
incommensurate case, consider initially taking the limit that the
penetration depth $\lambda\rightarrow \infty$. In this limit the
vector potential can be ignored,  and a Bezerinskii Kosterlitz
Thouless (BTK) transition, corresponding to the unbinding of
$\Phi_0/2$ defects, occurs at a transition temperature
${T_c}^*=\frac{\pi}{2}\rho_s$ \cite{BKT-1,BKT-2}. In reality,
since the penetration depth is finite, this will correspond to a
crossover temperature at which the resistivity starts to fall.
This is the case since superconducting vortices have a finite
energy and therefore exist at any finite temperature. Note that
this crossover temperature is half the value of the equivalent
BKT transition temperature for a conventional superconductor with
the same mean field $T_c$ and superfluid density. Consequently,
the corresponding  fluctuation regime will be larger. Now
consider the case in which there exist thermally excited unbound
superconducting vortex pairs. While such a state will not be
superconducting,  this does not imply that there is no longer any
type of order present (as is the case for conventional
superconductors). To see this, it is convenient to define
$\theta=(\phi_1+\phi_2)/2$ and $\phi=\phi_1-\phi_2$. The field
$\theta$ corresponds to conventional superconducting vortices and
is disordered. The field $\phi$ is uncoupled from the vector
potential and therefore has vortex-like defects that are not
screened. Vortices in this field correspond to dislocations in
the CDW order.  These dislocations cost a logarithmically
divergent energy and therefore are not present at sufficiently
low temperatures.  This implies that there can be quasi-long
range order in the CDW order parameter as defined through
Eq.~\ref{cdw}. The corresponding BKT transition at which the CDW
order is removed occurs at $T_{cdw}\le T_c^*$, above $T_{cdw}$,
strictly speaking, there is no order of any type.

The commensurate case is more interesting since, in principle,
it allows for the possibility to have $T_{cdw}>T^*$ (where $T^*$
corresponds to a crossover temperature at which the resistivity
starts to fall). To carry out a proper treatment, it is important
to include the following term in the free energy density(which
only exists in the commensurate case):
\begin{equation}
\epsilon[\Dpx^4(\Dnx^*)^4+\Dpy^4(\Dny^*)^4+\Dnx^4(\Dpx^*)^4+\Dny^4(\Dpy^*)^4].
\label{pert}
\end{equation}
In the London theory for Phase 3, this leads to an interaction
term $\frac{\epsilon|\psi|^8}{8}\cos(4\phi_1-4\phi_2)$. The
renormalization group (RG) equations describing the CDW
transition with such an interaction are the same as those derived
in Ref.~\cite{jos77}, in which two-dimensional $XY$ models
subject to clock model-like symmetry breaking fields are
considered.  These RG equations imply that $T_{cdw}$ for the CDW
order is  enhanced relative to that of the incommensurate case.
In the limit that Eq.~\ref{pert} can be treated as a
perturbation, we find that the enhancement of $T_{cdw}$ is given
by $T_{cdw}(h_4)=T_{cdw}(0)+\frac{\pi
e^{-\pi/16}}{16\sqrt{2}}|h_4|+O(h_4^2)$ where
$h_4=2a^2\epsilon|\psi|^8$ and $T_{cdw}(0)$ corresponds to the
BKT transition temperature with $h_4=0$.
%This theory
%also provides a means to further develop the related idea of a
%vortex disordered PDW phase that has been argued to be
%responsible for commensurate CDW order found in the pseudogap
%state of some cuprates \cite{che04}.

Finally, we turn to superconducting vortices containing a single
flux quantum.  These are the lowest energy vortices and they are
created by magnetic fields. The key point is that a single
$\Phi_0$ vortex can be either a conventional vortex or can be a
bound state of fractional flux defects. This possibility is
closely related to broken axisymmetric vortices discussed in the
context of superfluid $^3$He \cite{thu86,sal86}, unconventional
superconductors \cite{tok90,zhi95}, and in FFLO superconductors
\cite{agt07}. The dissociation of a conventional vortex into a
bound state of fractional flux defects provides a means to have a
vortex core structure with CDW order.  As a specific example, we
have determined the structure of a vortex in a superconductor
with non-vanishing $d$-wave and PDW superconductivity (the PDW
order is that of Phase 3 or 4).  There exists a solution in which
all components of the order parameter exhibit the same phase
winding. Furthermore, the $d$-wave order parameter vanishes at
the vortex core and the PDW order parameter is non-zero where the
$d$-wave order vanishes.  This vortex will have CDW order (at the
wavevectors listed in Table I) in the $d$-wave vortex core as
follows from Eq.~\ref{cdw}. Such a solution may be relevant for
understanding the observed CDW order inside the vortex cores of
some underdoped cuprates \cite{hof02}.

%Nevertheless, we point out an interesting phenomenological
%consequence of this picture. In particular, we often find that
%the stable vortex lattice structure is square when the
%conventional vortex disassociates. Of particular importance is
%that these square vortex lattice phases are stable, {\it even
%when the theory does not include the underlying crystal
%anisotropy} ($\kappa_2=0$). In other theories of tetragonal
%superconductors, the square vortex lattice is a consequence of
%the underlying crystal anisotropy \cite{kog97,fra97}. In this
%context, it is interesting to note that small angle neutron
%scattering measurements in NCCO have observed an orientationally
%disordered square vortex lattice; in which the local orientation
%of square vortex lattice is random with respect to the crystal
%lattice \cite{gil04}.

In summary, we have examined the broken symmetry phases of PDW
superconductors and have shown that the coexisting CDW or SDW
order provides a means to distinguish between these phases. We
have also shown that PDW superconductors exhibit topological
defects that include fractional superconducting vortices which
are coupled to dislocations in the coexisting CDW or SDW order.
These defects can play an important role in stabilizing
non-superconducting SDW or CDW phases. They are also important
in understanding the physics of superconducting single flux
quantum vortices where they can lead to CDW or SDW order inside
vortex cores. Finally, in the context of the cuprates,  we have
examined the competition between PDW and translational invariant
$d$-wave superconductivity and have shown that this prefers PDW
phases with intrinsic CDW order and that an additional SDW or CDW
appears due to the coexistence of $d$-wave and PDW
superconductivity.

Correspondence and requests for materials should be addressed to
DFA.

We acknowledge useful discussions with Kazumasa Hattori, Kazuo
Ueda, and Kun Yang.  D.F.A. and H.T. acknowledge hospitality from
the Kavli Institute for Theoretical Physics while part of this
work was done. This work was supported in part by the National
Science Foundation under Grant No. PHY05-51164.  H.T. was partly
supported by Scientific Research on Priority Areas Grants-in-Aid
(Nos. 19052003 and 17071011) from the MEXT of Japan.

%There is also an additional term in the commensurate case of the
%form
%\begin{eqnarray}
%\Dpx^8(\Dpy^*)^8+\Dpy^8(\Dnx^*)^8+\Dnx^8(\Dny^*)^8+\Dny^8(\Dpx^*)^8\ \+\Dpy^8(\Dpx^*)^8+\Dnx^8(\Dny^*)^8+\Dny^8(\Dpx^*)^8+\Dpx^8(\Dpy^*)^8
%\end{eqnarray}
%\begin{equation}
%F=\int dx^2 \{\frac{|\psi|^2}{4m}[(\vnabla \phi_1-e\vA)^2+(\vnabla \phi_3-e\vA)^2]+\frac{1}{2}\vB^2+\frac{\epsilon|\psi|^8}{8}\cos(4\phi_1-4\phi_3) \}\label{free2}
%\end{equation}
%\begin{equation}
%F=\int dx^2 \{\frac{|\psi|^2}{2m}(\vnabla \theta-e\vA)^2+\frac{1}{2}\vB^2+\frac{|\psi|^2}{8m}(\vnabla \phi)^2+\frac{\epsilon|\psi|^8}{8}\cos(4\phi)\}.
%\end{equation}

\end{document}